\documentclass[12pt,twoside]{article}
\usepackage{amssymb}
\usepackage{amsmath}
\usepackage{latexsym}
\usepackage{longtable}
\usepackage{epsfig}
\usepackage{graphicx,bbm,psfrag}
\graphicspath{{images/}}

\begin{document}
\begin{titlepage}
\vspace*{2cm}
\begin{center}
{\Large {\bf Energy Transport in Weakly Anharmonic Chains\bigskip\\}} 
{\large{Kenichiro Aoki$^\ast$, Jani Lukkarinen$^\star$, 
  Herbert Spohn$^\dag$}}\bigskip\bigskip\\
{$^\ast$Department of Physics, Keio University, \\4--1--1 Hiyoshi, Kouhoku--ku,
Yokohama 223--8521, Japan,\\
e-mail:~{\tt ken@phys-h.keio.ac.jp}\smallskip
}\\
{$^\star$Zentrum Mathematik, TU M\"unchen,
 D-85747 Garching, Germany,\\
 e-mail:~{\tt jlukkari@ma.tum.de}}\smallskip\\
{$^\dag$Zentrum Mathematik, TU M\"unchen,
 D-85747 Garching, Germany,\\
 e-mail:~{\tt spohn@ma.tum.de}}
\end{center}\bigskip\bigskip\bigskip
{\bf Abstract.}\medskip\\
We investigate the energy transport in a one-dimensional lattice
of oscillators with a harmonic nearest neighbor coupling and a
harmonic plus quartic on-site potential. As numerically observed
for particular coupling parameters 
before, and confirmed by our study, such chains satisfy Fourier's
law: a chain of length $N$ coupled to thermal reservoirs at both
ends has an average steady state energy current proportional to
$1/N$. On the theoretical level we employ the Peierls transport
equation for phonons and note that beyond a mere exchange of labels
it admits nondegenerate phonon collisions. These
collisions are responsible for a finite heat conductivity. The
predictions of kinetic theory are compared with molecular dynamics
simulations. In the range of weak anharmonicity, respectively 
low temperatures, reasonable agreement is observed.
\end{titlepage}
\newpage

\section{Introduction}\label{sec.1}
\setcounter{equation}{0}

In their seminal work of 1955, Fermi, Pasta, and Ulam \cite{FPU}
investigate the relaxation to 
equilibrium for a chain of coupled anharmonic oscillators,
by exploiting the then newly available
electronic computational devices. We refer to the informative
memorial volume \cite{1a}. Related to their study
is the issue of energy transport along the chain, which
is modelled by coupling a chain of length $N$ to thermal reservoirs
at both ends. With increased computational power at hand such
studies have been revived and carried through in considerable
detail. Excellent reviews are in print \cite{BLR00,LL03} and here we
only highlight, somewhat crudely, the
main findings:\medskip\\
(i) There are chains for which Fourier's law holds, in the sense
that the steady state energy flux $j_\mathrm{e}\cong
1/N$.\medskip\\
ii) There are chains with anomalous heat conduction, for which the
energy flux exhibits a power law dependence on $N$ which differs
from $1/N$.\medskip\\
iii) If the interaction depends only on the relative
displacements, anomalous heat conduction seems to be the
rule.\medskip\\
iv) In some models the conductivity depends on the details of the
coupling to the thermal reservoirs.\medskip

While the amount of data available is impressive, it is generally
agreed that there is very little theory which would serve as a
guideline. The harmonic chain can be solved exactly 
with the result  that $j_\mathrm{e}$ is independent of $N$
\cite{RLL67}. For the harmonic chain with random masses the transport for a
given wave 
number $k$ is proportional to $e^{-\gamma(k)N}$, $\gamma(k)>0$, with
$\gamma(k)\to 0$ as $k\to 0$. Thus the average energy current depends
on the precise spectral statistics of the thermal reservoir
\cite{CL74,KPW78}. For anharmonic chains there are attempts to
predict the exponent for the anomalous heat conduction through
mode-coupling theory \cite{NR02,LP98}.

In our note we follow the strategy of Peierls, who argues that in
case of weak nonlinearity one can use a Boltzmann type transport
equation for the computation of the thermal conductivity. For
anharmonic crystals in three dimensions phonon kinetic theory is
well supported through theory \cite{Zi,G86,S05} and also
experimentally \cite{Zi,TriH05}. Whether kinetic theory is
applicable to a weakly anharmonic chain is somewhat tentative.
There is no difficulty in writing down the appropriate transport
equation. Its collision term ensures energy and momentum
conservation in a phonon collision.
To progress further an analysis of the solution manifold to both
conservation laws becomes necessary. Firstly one has the trivial solutions
in which 
the two colliding phonons merely exchange their label. 
For the label exchanging solutions the collision operator vanishes.
However, as we first learned from Lefevere and Schenkel
\cite{LS05}, there is in addition a non-perturbative solution,
which leads to nondegenerate collisions. As will be explained in more
detail, with this input  kinetic theory predicts a finite, non-zero,
thermal conductivity even for a chain.

As we learned later on, Pereverzev \cite{P03} has already applied kinetic
theory to the FPU $\beta$-lattice, for which the potential energy
depends only on the relative displacements. For this model he
obtains the non-perturbative solution and argues,  based on
the linearized transport equation, that the energy current
correlation has a power law decay as $t^{-3/5}$. Thus the heat
transport is anomalous with $j_e\cong N^{2/5}$. In our
contribution we will be concerned with regular transport only.

To keep the model as simple as possible we will study the harmonic
chain with a small quartic on-site potential.  Equivalently, one may consider 
a fixed anharmonicity but ``low" temperatures. We regard the thermal
conductivity to be defined through the Green-Kubo formula and work
out the predictions of kinetic theory in fair detail. To
explicitly compute the conductivity one has to invert the
linearized collision operator. This is not completely
straightforward and we have to be satisfied with some estimates,
which however turn out to be sufficient for our purposes. The
predictions of kinetic theory will be compared with molecular
dynamics simulations. In fact, kinetic theory does rather well, with
a range of validity larger than expected on the basis of
purely theoretical arguments.


\section{The Green-Kubo formula, scaling properties}\label{sec.2}
\setcounter{equation}{0}

The anharmonic chain is governed by the Hamiltonian
\begin{equation}\label{2.1}
H=\sum_{j\in\mathbb{Z}}\{\tfrac{1}{2}p^2_j+\tfrac{1}{2}\omega^2_0q^2_j-
\delta\omega^2_0q_jq_{j+1}+\tfrac{1}{4}\lambda q^4_j\}\,.
\end{equation}
Here $q_j$ is the deviation from the rest position and $p_j$ the
momentum of the $j$-th particle. We choose units such that the
mass of a particle equals 1. $\omega_0$ and $\delta\omega_0$
characterize the harmonic on-site and nearest neighbor
interaction, respectively. $\omega_0>0$ and it has the dimensions of a
frequency. To have a stable harmonic part of $H$ we require
\begin{equation}\label{2.2}
0\leq\delta\leq\frac{1}{2}\,.
\end{equation}
In the border case $\delta=\frac{1}{2}$, the harmonic part can be
written as $\sum_j(\omega^2_0/4)(q_{j+1}-q_j)^2$ and thus depends
only on the relative displacement. $\lambda>0$ is the strength of
the quartic on-site potential.

The particular case $\lambda = 1$, $\delta = \frac{1}{2}$ is studied in
great detail in  \cite{AK02}, in which case kinetic theory is applicable at
low temperatures. For our  purposes it is of importance to add the extra
parameter $\delta$, since it is retained in the kinetic limit. Thereby one
can compare the theoretical predictions  with  
molecular dynamics simulations in their $\delta$-dependence.

To (\ref{2.1}) we associate the local energy
\begin{equation}\label{2.3}
H_j=\tfrac{1}{2}p^2_j+\tfrac{1}{2}\omega^2_0q^2_j+
\tfrac{1}{4}\lambda
q^4_j-\tfrac{1}{2}\delta\omega^2_0(q_{j-1}q_j+q_jq_{j+1})\,.
\end{equation}
Then
\begin{equation}\label{2.4}
\frac{d}{dt}H_j=-J_{j,j+1}+J_{j-1,j}
\end{equation}
with the energy current across the bond $j$ to $j+1$ given by
\begin{equation}\label{2.5}
J_{j,j+1}=-\tfrac{1}{2}\delta\omega^2_0(p_jq_{j+1}-p_{j+1}q_j)\,.
\end{equation}

The equilibrium state for (\ref{2.1}) is $Z^{-1}\exp[-H/T]$ with
$T$ denoting temperature. Equilibrium expectations are denoted by
$\langle\cdot\rangle^{(1/T)}$. We define the total energy current
correlation through
\begin{equation}\label{2.6}
C(t;T,\omega_0,\delta,\lambda)=\sum_{j\in\mathbb{Z}}\langle
J_{j,j+1}(0)J_{0,1}(t) \rangle^{(1/T)}\,.
\end{equation}
By stationarity of $\langle\cdot\rangle^{(\beta)}$, $C(t)=C(-t)$.
According to Green-Kubo the thermal conductivity at temperature
$T$ is defined by
\begin{equation}\label{2.7}
\kappa(T)= T^{-2}\int^\infty_0 dt\,C(t;T)\,.
\end{equation}
Regular transport in
the sense of Fourier's law requires that $0<\kappa(T)<\infty$.

$C(t)$ does not depend on all of its parameters separately. To find
the dependence out we transform to new variables as
\begin{equation}\label{2.8}
\tilde{q}_j(t)=\gamma q_j(\alpha t)\,,\;
\tilde{p}_j(t)=\alpha\gamma p_j(\alpha t)\,.
\end{equation}
Then $\tilde{\underline{q}}(t)$,
$\tilde{\underline{p}}(t)$ are solutions to Hamilton's
equations of motion for the Hamiltonian
\begin{equation}\label{2.9}
\widetilde{H}(\tilde{\underline{q}},\tilde{\underline{p}})=
\sum_{j\in\mathbb{Z}}\{\tfrac{1}{2}\tilde{p}^2_j+\tfrac{1}{2}
\alpha^2\omega^2_0\tilde{q}^2_j-\alpha^2\delta\omega^2_0\tilde{q}_j
\tilde{q}_{j+1}+\alpha^2\gamma^{-2}
\tfrac{1}{4}\lambda\tilde{q}^4_j\}\,.
\end{equation}
In addition,
\begin{equation}\label{2.9a}
H(\underline{q},\underline{p})=(\alpha\gamma)^{-2}
\widetilde{H}(\tilde{\underline{q}},\tilde{\underline{p}})\,.
\end{equation}
Therefore,
\begin{equation}\label{2.10}
C(t;T,\omega_0,\delta,\lambda)=(\alpha\gamma^2)^{-2}\alpha^{-4}C(t/\alpha;
\alpha^2\gamma^2T,\alpha\omega_0,\delta,\alpha^2\gamma^{-2}\lambda)\,,
\end{equation}
which by (\ref{2.7}) implies for the conductivity
\begin{equation}\label{2.11}
\kappa(T,\omega_0,\delta,\lambda)=\alpha^{-1}\kappa(\alpha^2\gamma^2T,
\alpha\omega_0,\delta,\alpha^2\gamma^{-2}\lambda)\,.
\end{equation}
Setting $\alpha\omega_0=1$, $\alpha^2\gamma^{-2}\lambda=1$ yields
the scaling form
\begin{equation}\label{2.12}
\kappa(T,\omega_0,\delta,\lambda)=\omega_0 \Xi(\omega^{-4}_0\lambda
T,\delta) \,.
\end{equation}

In our molecular dynamics simulations, see Sect.~\ref{sec:MD}, we
set $\omega_0=1/\sqrt{\delta}$ and $\lambda=1$. On the other hand,
for the kinetic limit the natural choice is $\omega_0=1$ and fixed
inverse temperature $\beta$. Inserting 
$\alpha=\sqrt{\delta}$, $\gamma=(\delta\beta T)^{-1/2}$ in (\ref{2.11}) one
arrives at 
\begin{equation}\label{2.13}
T^2\kappa(T,\frac{1}{\sqrt{\delta}},\delta,1)=
\frac{1}{\sqrt{\delta}}\frac{1}{\delta^4\beta^2}(\delta^2\beta
T)^2\kappa(\beta^{-1},1,\delta,\delta^2\beta T)\,.
\end{equation}
Thus the limit $T\to 0$ on the left corresponds to the limit
$\lambda\to 0$ of $\lambda^2\kappa(\beta^{-1},1,\delta,\lambda)$ on the
right hand side, which will be studied in the following section.


\section{Energy current-current correlation in the kinetic limit}\label{sec.3}
\setcounter{equation}{0}

In this section we set $\omega_0=1$. To study the kinetic limit it
is convenient to switch to Fourier space. Let
$\mathbb{T}=[-\frac{1}{2},\frac{1}{2}]$ be the first Brillouin zone
of the lattice dual to $\mathbb{Z}$. For $f:\mathbb{Z}\to
\mathbb{R}$ we set
\begin{equation}\label{3.1}
\widehat{f}(k)=\sum_{j\in\mathbb{Z}}e^{-i2\pi kj}f_j\,,\quad
k\in\mathbb{T}\,,
\end{equation}
with the inverse
\begin{equation}\label{3.2}
f_j=\int_\mathbb{T} dk e^{i2\pi kj}\widehat{f}(k)\,.
\end{equation}

We decompose
\begin{equation}\label{3.3}
H=H_\mathrm{har}+\tfrac{1}{4}\lambda\sum_{j\in \mathbb{Z}}
q_j^4\,.
\end{equation}
The harmonic part $H_\mathrm{har}$ has the dispersion relation
\begin{equation}\label{3.4}
\omega(k)=\big(1-2\delta\cos(2\pi k)\big)^{1/2}\,.
\end{equation}
$p_j,q_j$, $j\in \mathbb{Z}$, are concatenated into a single
complex-valued field through
\begin{equation}\label{3.5}
a(k)=\frac{1}{\sqrt{2}}\big(\sqrt{\omega(k)}\widehat{q}(k)+
i\frac{1}{\sqrt{\omega(k)}}\widehat{p}(k)\big)\,.
\end{equation}
Notationally it will be convenient to also define  
\begin{equation}\label{3.6}
a(k)^\ast=a(k,1)\,,\quad a(k)=a(k,-1)\,.
\end{equation}
Then the equations of motion read
\begin{eqnarray}\label{3.7}
&&\hspace{-45pt} \frac{d}{dt}a(k,t)= -i\omega(k)a(k,t)
-i\lambda\sum_{\sigma_1,\sigma_2,\sigma_3=\pm1}
\int_{\mathbb{T}^3}dk_1 dk_2 dk_3\nonumber\\
&&\hspace{-10pt}(16\omega(k)\omega(k_1)\omega(k_2)\omega(k_3))^{-1/2}
\delta(k+\sigma_1k_1+\sigma_2k_2+\sigma_3k_3)\prod^3_{j=1}
a(k_j,\sigma_j,t)\,.
\end{eqnarray}

In these new variables the harmonic part of the energy becomes
\begin{equation}\label{3.8}
H_\mathrm{har}=\int_\mathbb{T} dk \omega(k)a(k)^\ast a(k)
\end{equation}
and the total current, $J_\textrm{tot}=\sum_j J_{j,j+1}$, becomes
\begin{equation}\label{3.9}
J_\textrm{tot}=\delta\int_\mathbb{T}dk \sin(2\pi k)a(k)^\ast a(k)=
\frac{1}{2\pi}\int_\mathbb{T}dk(\frac{d}{dk}\omega(k))\omega(k)a(k)^\ast
a(k)\,.
\end{equation}

We plan to study the energy current correlation (\ref{2.6}) in the
limit of small $\lambda$. Since in equilibrium $\langle
J_{0,1}(t)\rangle^{(\beta)}=0$, it holds
\begin{equation}\label{3.10}
\sum_{j\in\mathbb{Z}}\langle J_{0,1}(t)J_{j,j+1}\rangle^{(\beta)} =
\lim_{\tau\to 0}\frac{1}{\tau}\langle J_{0,1}(t)\rangle^{(\beta,\tau)}\,,
\end{equation}
where $\langle\cdot\rangle^{(\beta,\tau)}$ refers to expectation with
respect to the perturbed equilibrium measure $Z^{-1}\exp[-\beta H+
\tau J_\textrm{tot}]$, $J_\textrm{tot}$ the total current, in the
infinite volume limit. For small $\lambda$ we can ignore the
quartic on-site potential in the average
$\langle\cdot\rangle^{(\beta,\tau)}$. Thus
$\langle\cdot\rangle^{(\beta,\tau)}$ is replaced by the Gaussian
measure $\langle\cdot\rangle^{\beta,\tau}_{\mathrm{harm}}$, which is uniquely
characterized through its covariance
\begin{eqnarray}\label{3.11}
&&\hspace{-30pt}\langle a(k)^\ast a(k')\rangle^{\beta,\tau}_{\mathrm{harm}}=
\delta(k-k')\big(\beta\omega(k)-\tau\delta\sin(2\pi
k)\big)^{-1}\,,\nonumber\\
&&\hspace{-30pt}\langle a(k)\rangle^{\beta,\tau}_{\mathrm{harm}} =0\,,\; \langle
a(k) a(k')\rangle^{\beta,\tau}_{\mathrm{harm}}=0\,.
\end{eqnarray}
The state $\langle \cdot\rangle^{\beta,\tau}_{\mathrm{harm}}$ is not invariant 
under the mechanical time evolution and we have to understand how for 
small $\lambda$ such a
non equilibrium measure evolves in time. Instead of 
$\langle \cdot\rangle^{\beta,\tau}_{\mathrm{harm}}$ let us consider as
initial state an 
arbitrary translation invariant Gaussian measure with covariance
\begin{eqnarray}\label{3.12}
&&\hspace{-30pt}\langle a(k)\rangle_0= 0\,,\quad \langle a(k)
a(k')\rangle_0=0\,,\nonumber\\
&&\hspace{-30pt}\langle a(k)^\ast a(k')\rangle_0 =
\delta(k-k')W(k)\,,
\end{eqnarray}
compare with (\ref{3.11}). Its two-point function at time $t$ is
given by
\begin{equation}\label{3.13}
\langle a(k,t)^\ast a(k',t)\rangle_0 = \langle a(k)^\ast
a(k')\rangle_t = \delta(k-k')W^\lambda(k,t)\,,
\end{equation}
which defines $W^\lambda(k,t)$. In particular, by (\ref{3.9}),
\begin{equation}\label{3.13a}
\langle J_{0,1}(t)\rangle_0=\delta\int_\mathbb{T}dk\sin(2\pi
k)W^\lambda(k,t)\,.
\end{equation}
For $\lambda=0$, one has $W^{\lambda=0}(k,t)=W(k)$. For small
$\lambda$ the time variation of $W^\lambda(k,t)$ is on the
time-scale $\lambda^{-2}$. Thus one expects that the following
limit exists,
\begin{equation}\label{3.14}
\lim_{\lambda\to 0}W^\lambda(k,\lambda^{-2}t)=W(k,t)\,,
\end{equation}
and that the limit Wigner function $W(k,t)$ evolves according to
the spatially homogeneous Boltzmann equation
\begin{eqnarray}\label{3.15}
&&\hspace{-35pt}\frac{\partial}{\partial
t}W(k,t)=12\pi\sum_{\sigma_1,\sigma_2,\sigma_3=\pm1}
\int_{\mathbb{T}^3}dk_1dk_2dk_3(16\omega\omega_1\omega_2\omega_3)^{-1}
\nonumber\\
&&\hspace{30pt}\delta(\omega+\sigma_1\omega_1+\sigma_2\omega_2+\sigma_3\omega_3)
\delta(k+\sigma_1k_1+\sigma_2k_2+\sigma_3k_3)\nonumber\\
&&\hspace{30pt}\big[W_1W_2W_3+
W(\sigma_1W_2W_3+\sigma_2W_1W_3+\sigma_3W_1W_2)\big]
\end{eqnarray}
with initial conditions $W(k,0)=W(k)$. Here, and later on, we use
the shorthand $\omega=\omega(k)$, $\omega_j=\omega(k_j)$,
$W=W(k)$, $W_j=W(k_j)$, $j=1,2,3$. In Appendix A we explain the
second order diagrammatic expansion in $\lambda$, mostly to make
sure that the collision strength is correct.

By the argument in Appendix 18.1 of \cite{S05}, energy and
momentum conservation in (\ref{3.15}) can be satisfied only if
\begin{equation}\label{3.16}
  1+\sum^3_{j=1}\sigma_j=0\,,
\end{equation}
i.e., only for phonon number conserving collisions. Hence
(\ref{3.15}) simplifies to
\begin{eqnarray}\label{3.17}
&&\hspace{-24pt}\frac{\partial}{\partial t}W(k,t)=\frac{9\pi}{4}
\int_{\mathbb{T}^3}dk_1dk_2dk_3(\omega\omega_1\omega_2\omega_3)^{-1}
\delta(\omega+\omega_1-\omega_2-\omega_3)
\delta(k+k_1-k_2-k_3)\nonumber\\
&&\hspace{42pt}\big[W_1W_2W_3+
W(-W_1W_2-W_1W_3+W_2W_3)\big]\nonumber\\
 &&\hspace{30pt}=\mathcal{C}\big(W(t)\big)(k)\,.
\end{eqnarray}
Clearly, the equilibrium Wigner function
\begin{equation}\label{3.18}
W_\beta(k)=\frac{1}{\beta\omega(k)}
\end{equation}
is a stationary solution for (\ref{3.17}). Since according to
(\ref{3.10}) and (\ref{3.11}) only small deviations from equilibrium
are needed, we linearize the collision operator as
\begin{equation}\label{3.19}
\mathcal{C}(W_\beta+(W_\beta)^2f)=-\beta^{-4}
Lf+\mathcal{O}(f^2)\,.
\end{equation}
Then
\begin{eqnarray}\label{3.20}
&&\hspace{-24pt}Lf(k)=\frac{9\pi}{4}
\int_{\mathbb{T}^3}dk_1dk_2dk_3(\omega\omega_1\omega_2\omega_3)^{-2}
\delta(\omega+\omega_1-\omega_2-\omega_3)\nonumber\\
&&\hspace{24pt}
\delta(k+k_1-k_2-k_3)\big(f(k)+f(k_1)-f(k_2)-f(k_3)\big)\,.
\end{eqnarray}
For this particular choice of linearization $L=L^\ast$ in
$L^2(\mathbb{T},dk)$.

Let $Vf(k)=W_\beta(k)f(k)$. Then
\begin{equation}\label{3.20a}
  \mathcal{C}(W_\beta+f)= - Af+\mathcal{O}(f^2)=
  -\beta^{-4}LV^{-2}f+\mathcal{O}(f^2)\,.
\end{equation}
Let $g(k)=\sin(2\pi k)$ and let $\langle\cdot,\cdot\rangle$ be the
scalar product in $L^2(\mathbb{T},dk)$. Then, with $W_\tau(k,t)$
denoting the solution to (\ref{3.17}) for the initial condition
$W(k)=\big(\beta\omega(k)-\tau\delta\sin(2\pi k)\big)^{-1}$, one
has
\begin{eqnarray}\label{3.21}
&&\hspace{-80pt}\lim_{\lambda\to
0}C(\lambda^{-2}t;\beta^{-1},1,\delta,\lambda)=\lim_{\tau\to 0}
\frac{1}{\tau}\delta\langle g,W_\tau(t)\rangle\nonumber\\
&&\hspace{42pt}=\delta^2\langle g,e^{-At}V^2g\rangle\nonumber\\
&&\hspace{42pt}=\delta^2\langle
g,Ve^{-\beta^{-4}V^{-1}LV^{-1}t}Vg\rangle\,.
\end{eqnarray}
Integrating over time one concludes that
\begin{eqnarray}\label{3.22}
&&\hspace{-80pt}\lim_{\lambda\to
0}\lambda^2\beta^{-2}\kappa(\beta^{-1},1,\delta,\lambda)=
\delta^2\beta^4\langle g,V(V^{-1}LV^{-1})^{-1}Vg\rangle\nonumber\\
&&\hspace{42pt}=\delta^2\beta^4\langle g,V^2L^{-1}V^2g\rangle\nonumber\\
&&\hspace{42pt}=\delta^2\langle\omega^{-2}g,L^{-1}\omega^{-2}g\rangle\,.
\end{eqnarray}
Combining with (\ref{2.13}) yields as the final result
\begin{equation}\label{3.23}
\lim_{T\to 0}T^2\kappa(T,\frac{1}{\sqrt{\delta}},\delta,1)=
\delta^{-5/2}\langle\omega^{-2}g,L^{-1}\omega^{-2}g\rangle\,.
\end{equation}
The inner product on the right depends only on $\delta$. The
prefactor is slightly misleading, since $L$ is proportional to
$\delta^{-1}$. Hence
\begin{equation}\label{3.23a}
\delta^{-5/2}\langle\omega^{-2}g,L^{-1}\omega^{-2}g\rangle
=\delta^{-3/2}c(\delta)
\end{equation}
with $0<c(\delta)<\infty$.

(\ref{3.22}) can rephrased as the asymptotics of the scaling
function from (\ref{2.12}),
\begin{equation}\label{3.24}
\lim_{x\to 0}x^{2}\Xi(x,\delta)=\delta^2\langle\omega^{-2}g,L^{-1}
\omega^{-2}g\rangle\,.
\end{equation}
In other words
\begin{equation}\label{3.25}
\kappa(T,\omega_0,\delta,\lambda)\cong (\omega_0)^9(\lambda
T)^{-2}\delta^2\langle\omega^{-2}g,L^{-1} \omega^{-2}g\rangle
\end{equation}
for small $\omega^{-4}_0\lambda T$. Our argument provides no
indication over which range (\ref{3.25}) is a valid approximation.
In fact, even the claim (\ref{3.23}), (\ref{3.23a}) is tentative.
The diagrammatic expansion from Appendix A relies on the
separation into leading and subleading diagrams. For a
three-dimensional lattice such a separation is convincing and can
be checked for special diagrams. The oscillatory time integrals
for the chain have a slower decay and the rough estimates used so far 
are not sufficient to justify the separation into leading and subleading
diagrams, which we have assumed here. On the other hand, there could very
well be 
cancellations which are difficult to access through the expansion
in Feynman diagrams. Also we need the validity of the kinetic
equation only close to thermal equilibrium. In view of this
situation a molecular dynamics simulation is in demand.


\section{Thermal conductivity from the one-dimensional Boltzmann equation}\label{sec.4}
\setcounter{equation}{0}

\begin{figure}
  \begin{center}
    \includegraphics*[width=0.9\textwidth]{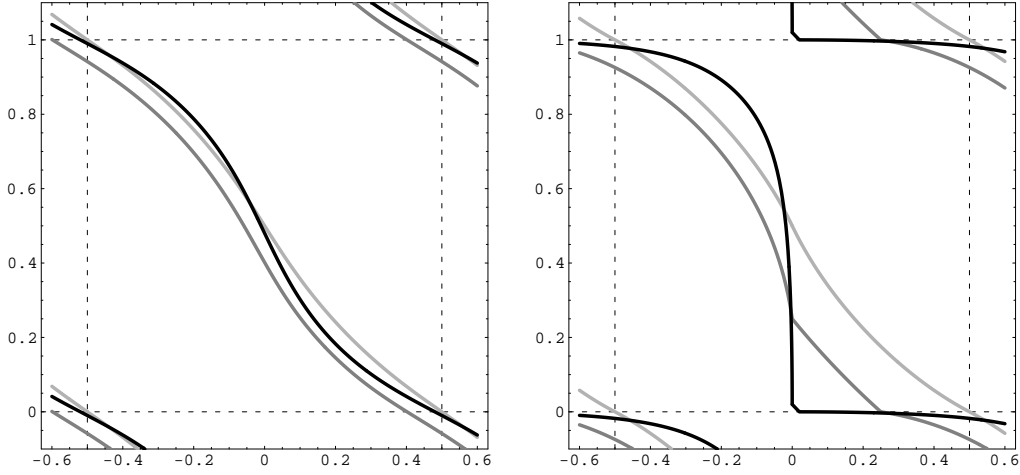}
    \caption{The non-perturbative solution $h(k_1;k_3,\delta)$
      as a function of $k_1$ for fixed $k_3$ and $\delta$.
      On the left, $\delta=0.4$, and on the right, $\delta=0.5$.
      For each value three solutions are plotted, corresponding to
      $k_3=0.02$ (black), $0.25$ (dark grey), and $0.5$ (light grey).  We plot
      $\mathbb{T}\times \mathbb{T}$ in the extended
      zone scheme, the dashed lines are the
      boundaries of a unit cell. For $\delta < 0.5$ the non-perturbative
      solution is 
smooth, while for  $\delta = 0.5$ there are cusp singularities.}
    \label{fig:umklapps}
  \end{center}
\end{figure}

The linearized collision operator $L$ of (\ref{3.20}) corresponds to the
quadratic form
\begin{eqnarray}\label{4.1}
&&\hspace{-24pt}\langle f,Lf\rangle=\frac{1}{4}\frac{9\pi}{4}
\int_{\mathbb{T}^4}dk_1dk_2dk_3dk_4(\omega_1\omega_2\omega_3\omega_4)^{-2}
\delta(\omega_1+\omega_2-\omega_3-\omega_4)\nonumber\\
&&\hspace{28pt} \delta(k_1+k_2-k_3-k_4)(f_1+f_2-f_3-f_4)^2\,.
\end{eqnarray}
We use momentum conservation to integrate over $k_4$. For energy
conservation we thus need the solutions to
\begin{equation}\label{4.2}
\omega(k_1)+\omega(k_2)-\omega(k_3)-\omega(k_1+k_2-k_3)=0\,.
\end{equation}
The obvious solutions are
\begin{equation}\label{4.3}
k_1=k_3\quad\textrm{and}\quad k_2=k_3\,.
\end{equation}

Expanding (\ref{4.2}) to first order in  $\delta$, as noticed in
\cite{LS05},  there  
is yet a  further, {\it non-perturbative} solution
given by
\begin{equation}\label{4.4}
k_1+k_2=\frac{1}{2}\quad \textrm{modulo 1 for all }k_3\,.
\end{equation}
This suggests to write, in general,
\begin{equation}\label{4.5}
k_2=h(k_1;k_3)\,.
\end{equation}
Indeed, for every $k_3\in\mathbb{T}$ there exists a unique
function $h(\cdot;k_3):\mathbb{T}\to\mathbb{T}$, which is
continuous, one-to-one, and satisfies
\begin{equation}\label{4.6}
\omega(k_1)+\omega(h(k_1;k_3))-\omega(k_3)-\omega(k_1+h(k_1;k_3)-k_3)=0\,.
\end{equation}
$h$ is called the non-perturbative solution. In Fig.~\ref{fig:umklapps}
we display a few non-perturbative solutions at
$\delta=0.4$ and $\delta=0.5$ for three values of $k_3$.
For small $\delta$ one finds
\begin{equation}\label{4.7}
h(k_1;k_3)=\frac{1}{2}-k_1-\delta\frac{1}{2\pi}\big(\sin(2\pi
k_1)+\sin(2\pi k_3)\big)+\mathcal{O}(\delta^2)\,,
\end{equation}
which reasonably well approximates the left hand of Fig.~\ref{fig:umklapps}.

Inserting the solutions (\ref{4.3}) and (\ref{4.5}) to energy conservation
into (\ref{4.1}) 
splits the linearized collision operator as the sum
\begin{equation}\label{4.7a}
L = L_{\mathrm{ex}}+L_{\mathrm{npert}}\,.
\end{equation}
By symmetry,  for the
label exchanging solution 
$\langle f,L_{\mathrm{ex}}f\rangle=0$ for all $f$.
Hence $L_{\mathrm{ex}}=0$ and energy conservation in (\ref{4.1}) will be
evaluated always at $k_2=h(k_1;k_3)$.

In phonon kinetic theory it is customary to distinguish between
normal and umklapp processes. We choose the convention that
$k_j\in[-\frac{1}{2},\frac{1}{2}]$, $j=1,\ldots,4$. Then a process
is called normal if $k_1+k_2-k_3-k_4=0$, while it is umklapp if
$k_1+k_2-k_3-k_4=\pm 1$. By this definition, the curves in
Fig.~\ref{fig:umklapps} are divided into a normal piece and an umklapp
piece. For example, at $\delta=0$, one has $k_1+k_2=\pm 1/2$ and
$k_3+k_4=\pm 1/2$. If the two terms have opposite sign, the
collision process is normal and otherwise it is umklapp. In our
context such a division looks artificial. In fact we will find
that both, normal and umklapp, processes contribute to the thermal
conductivity.

We turn to the zero subspace of $L$, i.e., to solutions of $Lf=0$
in $L^2(\mathbb{T})$. By (\ref{4.1}), clearly they must be
collisional invariants in the sense that
\begin{equation}\label{4.8}
f(k_1)+f(h(k_1;k_3))-f(k_3)-f(k_1+h(k_1;k_3)-k_3)=0
\end{equation}
for all $(k_1,k_3)\in \mathbb{T}^2$. The obvious solutions are
\begin{equation}\label{4.9}
f(k)=1\,,\quad f(k)=\omega(k)\,.
\end{equation}
We expect that there are no further solutions, but no proof is
available, at present. This is an important issue, since speaking
in general, the number of collisional invariants is the crucial
information on the long-time behavior of a kinetic equation. At
$\delta=0$, $h$ does not depend on $k_3$ and as a consequence the
zero subspace of $L$ becomes infinite-dimensional consisting of
all $f$'s satisfying $f(k_1)+f(\frac{1}{2}-k_1)=0$.

We integrate in (\ref{4.1}) over $k_4$ and $k_2$. For the volume
element with respect to $k_2$ we need
\begin{eqnarray}\label{4.10}
&&\hspace{-10pt}\frac{\partial}{\partial
k_2}\big(\omega(k_1)+\omega(k_2)-\omega(k_3)-
\omega(k_1+k_2-k_3)\big)\big|_{k_2=h(k_1;k_3)}\\
&&\hspace{0pt} =(2\pi\delta)\big(\omega(k_2)^{-1}
\sin(2\pi k_2)-\omega(k_1+k_2-k_3)^{-1}
\sin(2\pi(k_1+k_2-k_3))\big)\big|_{k_2=h(k_1;k_3)}\,.\nonumber 
\end{eqnarray}
Hence
\begin{eqnarray}\label{4.11}
&&\hspace{-26pt}\langle
f,Lf\rangle=\frac{9\pi}{16}(2\pi\delta)^{-1}
\int_{\mathbb{T}^2}dk_1dk_3\big(\omega(k_1)\omega(k_2)\omega(k_3)
\omega(k_1+k_2-k_3)\big)^{-2}\nonumber\\
&&\hspace{28pt} |\omega(k_2)^{-1}\sin(2\pi
k_2)-\omega(k_1+k_2-k_3)^{-1}\sin(2\pi(k_1+k_2-k_3))|^{-1}\nonumber\\
&&\hspace{28pt}
\big(f(k_1)+f(k_2)-f(k_3)-f(k_1+k_2-k_3)\big)^2\big|_{k_2=h(k_1;k_3)}
\end{eqnarray}
and $L$ carries an explicit prefactor $\delta^{-1}$, as claimed in
(\ref{3.23a}).

To obtain the thermal conductivity (in the kinetic regime) one has
to invert $L$, which can be achieved  only numerically and which is
not completely straightforward because of the constraint due to energy
conservation. However for  
small $\delta$, say up to $\delta = 0.35$, more modest means already suffice.
By Jensen's
inequality one has 
\begin{equation}\label{4.12}
\langle f,L^{-1}f\rangle\geq \langle f,f\rangle^2/\langle
f,Lf\rangle\,
\end{equation}
with $f(k) = \omega(k)^{-2}\sin(2\pi k)$. For $\delta=0$ one obtains
\begin{equation}\label{4.13}
\langle f,f\rangle^2=\big(\int^{1/2}_{-1/2}(\sin 2\pi
k)^2dk\big)^2=\frac{1}{4}
\end{equation}
and
\begin{eqnarray}\label{4.14}
&&\hspace{-20pt}\langle f,Lf\rangle=\delta^{-1} \frac{9}{32}
\int^{1/2}_{-1/2}dk_1\int^{1/2}_{-1/2}dk_3|\sin (2\pi k_1)-\sin
(2\pi k_3)|^{-1}\nonumber\\
&&\hspace{36pt}\big(2\sin (2\pi k_1)-2\sin (2\pi k_3)\big)^2\nonumber\\
&&\hspace{24pt}=\delta^{-1}\frac{9}{\pi^2}\,.
\end{eqnarray}
Combing (\ref{3.23}) and (\ref{4.12}), (\ref{4.13}), (\ref{4.14}) yields
\begin{equation}\label{4.15}
\lim_{T\to 0}T^2\kappa(T,\frac{1}{\sqrt{\delta}},\delta,1)\geq
\delta^{-3/2}\frac{\pi^2}{36}\cong\delta^{-3/2}\,0.27
\end{equation}
 for small
$\delta$.

For numerical inversion of $L$ at $\delta=0$ we expand in a basis
of the form $\sin((2n+1)2\pi k)$, $n=0,1,\ldots$, and, instead of $c(0) = 0.27$,  obtain
 the prefactor in (\ref{3.23a}) as $c(0) =  0.275637$  with a
stable value over the range
$n=5,\ldots,32$. As can be seen from Fig. \ref{fig:umklapps},
even at $\delta = 0.4$ the approximate solution (\ref{4.7}) is rather accurate.
Therefore $c(\delta)$ is expected to depend only weakly on $\delta$.
Our claim is supported by the lower bound
\begin{equation}\label{4.16}
c(\delta) = \delta^{-1} \langle\omega^{-2}g,L^{-1}\omega^{-2}g \rangle \geq  
\langle\omega^{-3/2}g,\omega^{-3/2}g \rangle^2\big/\delta  \langle
\omega^{-1}g,L\omega^{-1}g \rangle\,,
\end{equation}
$g(k) = \sin(2\pi k)$. For this particular choice of the variational
function the singular denominator in (\ref{4.11}) 
is cancelled exactly and the numerical integration, using the true
non-perturbative solution, becomes routine.  
The lower bound to 
$c(\delta)$ drops from $0.27$ at $\delta = 0$
to $0.2$ at $\delta = 0.35$.
More details on the linearized collision operator can be found in \cite{LS06}.


\section{Thermal conductivity from molecular dynamics}
\label{sec:MD}
\setcounter{equation}{0}

\begin{figure}
  \centering
  \includegraphics*[width=0.8\textwidth]{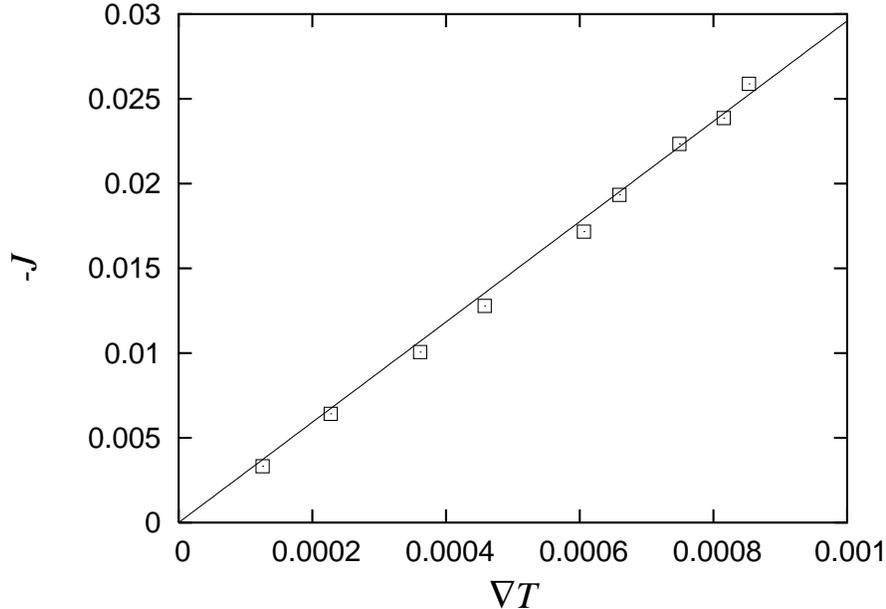}
  \caption{The relation between the average current $J/N$ and the
    temperature gradient, $\nabla T$, for $T=0.4, \delta=0.2,
    N=200$. The straight line is Fourier's law as obtained by a fit
    with gradient $\kappa$.}
  \label{fig:fourier}
\end{figure}

Kinetic theory is expected to be valid for small dimensionless
coupling $\omega_0^{-4}\lambda T$ and large system size. How small a coupling
and how large a system size can be explored only through molecular dynamics
simulations. To this end, we compute thermal conductivities for
various parameters of the anharmonic chain (\ref{2.1}). We set $\lambda = 1$,
$\omega_0 = 1/\sqrt{\delta}$. Then the infinite volume  
conductivity depends only on $\delta,T$ and kinetic theory becomes valid in
the limit $T \to 0$, compare with Sects. \ref{sec.2} and \ref{sec.3}. In
particular, 
\begin{equation}\label{5.0}
T^2\kappa(T) \simeq
\delta^{-3/2}c(\delta)
\end{equation}
for small $T$ with $c(0) = 0.28$ and $c(\delta)$ slowly dropping to smaller
values  
as $\delta$ is increased.

To numerically determine the thermal conductivity we take a chain of finite
length, $N$,  and attach thermostats at both ends. We use 
free boundary conditions, which means $q_{j+1}=q_j$ at the boundaries,
but the
physical results are insensitive to this particular choice of
boundary conditions. We adopt deterministic thermostats which
generalize those of Nos\'e--Hoover \cite{NH84,NH84b}, and
follow the methods used for the chain when $\delta=1/2$
\cite{AK02,AK00}. The non--equilibrium steady state is achieved by
integrating the equations of motion and physical observables are
measured by averaging over time after waiting for a sufficiently 
long equilibration time span. The integration is performed
numerically using standard algorithms, such as Runge--Kutta, with
time steps of $0.001\sim0.04$. $10^7\sim10^{10}$ samples are taken
to obtain the average values of physical observables. The results
do not depend on the time step size. A non--equilibrium steady
state has as its parameters
 the boundary temperatures, $N$, and $\delta$. The local temperature is
defined through the relation $T_j=\langle p_j^2\rangle$.

In principle, a single non--equilibrium state suffices to obtain
$\kappa$. To determine $\kappa$ with enough accuracy we employ a
more elaborate procedure. We choose
boundary conditions with varying temperature gradients, $\nabla T$, but
with the 
same temperature close to the midpoint. We can then check that the
average current is linear in $\nabla T$ and obtain $\kappa$ for given
$T,\delta$, and $N$, as illustrated in
Fig.~\ref{fig:fourier}. The temperature gradient needs to be
computed away from the boundaries because of jumps in the 
temperature at both ends, see Sect.\;{\ref{sec.5} for a discussion.
Since local energy is conserved and there
are no internal heat sinks or sources, the average current is constant
throughout the system. We increase $N$
and examine if the bulk limit is reached to finally extract
$\kappa$ for given $T,\delta$. Bulk behavior had already been observed
when $\delta=1/2$ \cite{AK00,HZ00}, and we do so also in the present study.

\begin{figure}
  \centering
  \includegraphics*[width=0.48\textwidth]{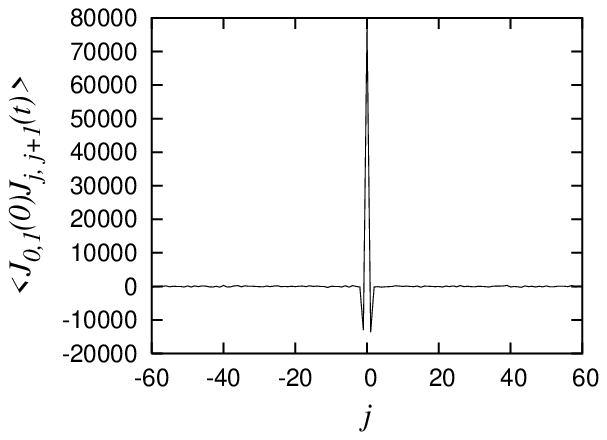}
  \includegraphics*[width=0.48\textwidth]{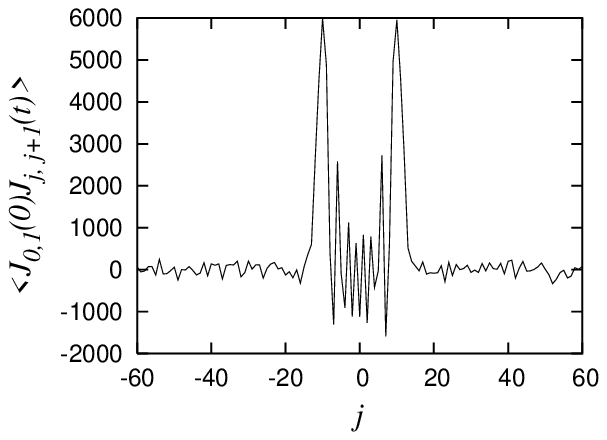}\\
  \includegraphics*[width=0.48\textwidth]{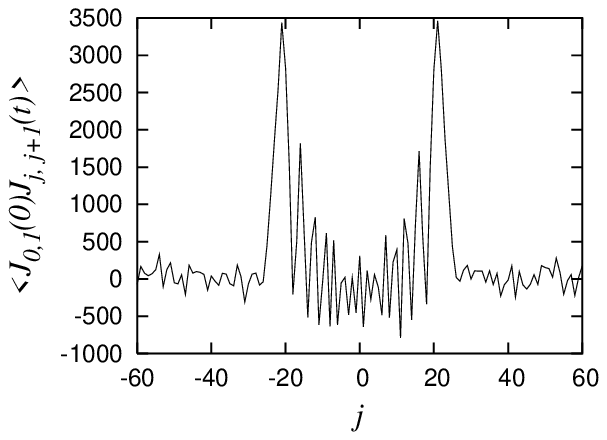}
  \includegraphics*[width=0.48\textwidth]{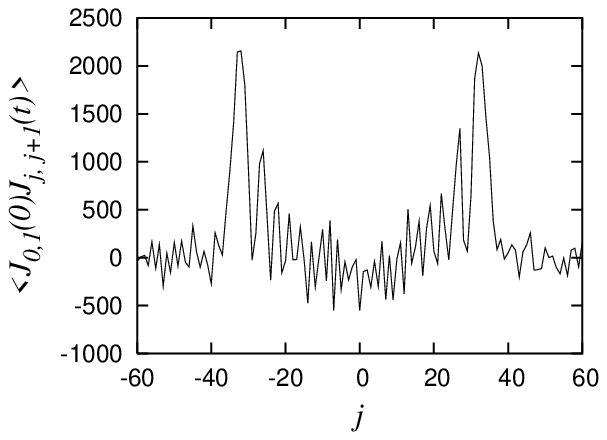}
  \caption{The autocorrelation function $\langle
  J_{0,1}(0)J_{j,j+1}(t)\rangle$ at $t=0,20,40,60$ for $T=0.4,
  \delta=1/3$. The 
  peaks move away at a constant rate, from which the speed of phonons
  is extracted.  }
  \label{fig:JJ}
\end{figure}

Naively, it would seem that weak coupling physics should be easy to
simulate. 
The computational difficulties arise because the mean
free path inevitably becomes large and the relaxation towards the
steady state becomes slow, which demands to carry out the
simulation for large system sizes and sufficiently long times.
These factors limit the range of accessible parameters values. 
To elucidate these points and also to understand the kinetic
theory aspects of this system better, we analyze the statistical
properties of the system in more detail. To 
judge the required system size we have to estimate the mean free
path $\Lambda$. It is not a sharply defined quantity. Following
Ziman \cite{Zi} we set
\begin{equation}\label{MD.1}
\Lambda=\kappa/(C_v \overline{v})\,,
\end{equation}
where $C_v$ is the specific heat and $\overline{v}$ the average
speed of phonons.
In the harmonic approximation one obtains
$C_v=1$. Kinetically the equilibrium phonon number density equals
$T/\omega(k)$, which suggests to set
\begin{equation}\label{MD.2}
\overline{v}=\Big(\int_{\mathbb{T}}
(\frac{\omega'}{2\pi})^2
T\omega^{-1}dk\Big/\int_{\mathbb{T}}T\omega^{-1}dk\Big)^{1/2}
\end{equation}
with the following expansion for small $\delta$,
\begin{equation}\label{MD.2a}
\overline{v}= \sqrt{\frac{\delta}{2}}
\left(1+\frac{9}{16}\delta^2+{\cal O}(\delta^4)\right)\,.
\end{equation}

\begin{figure}
  \centering
  \includegraphics*[width=0.8\textwidth]{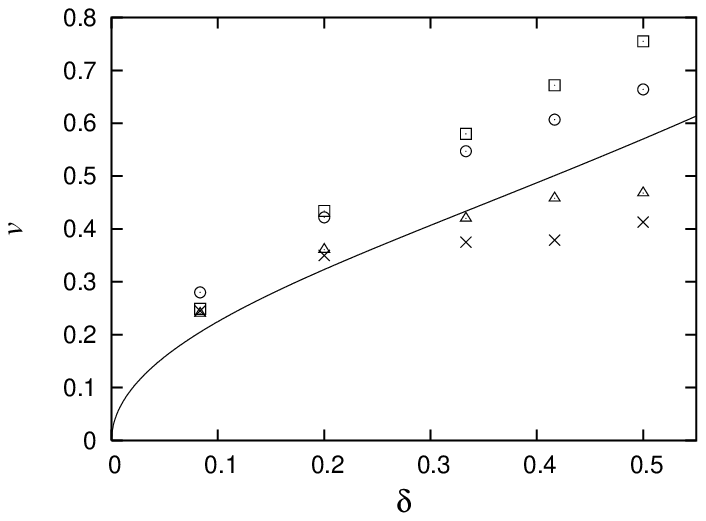}
  \caption{Average phonon velocity, $\overline v$, for 
    $T=0.1\ (\Box), \ 0.4\ (\bigcirc),\ 4 (\bigtriangleup),\ 10
    (\times)$.  
    The solid line represents the kinetic approximation (\ref{MD.2a}). }
  \label{fig:vPhonon}
\end{figure}

The speed of phonon propagation can also be measured directly from
the time and space dependences of the autocorrelation function
$\langle J_{0,1}(0)J_{j,j+1}(t)\rangle$ in thermal equilibrium
\cite{LP98,AK02}.   The velocity of the peaks  in the correlation
function is equated with the average phonon velocity relevant for thermal
transport. An example is shown in Fig.~\ref{fig:JJ}.  The measured
velocities can be compared to the kinetic result in
(\ref{MD.2a}), which is done in Fig.~\ref{fig:vPhonon}.  Perfect
agreement is not expected for a number of reasons. First, there is no
unique definition of the average speed of phonons so that there is
no guarantee that the average  (\ref{MD.2}) is
precisely what we measure in the molecular dynamics simulations as in
Fig.~\ref{fig:JJ}. Furthermore, the kinetic approximation for
$\overline v$ does not include the effects of anharmonicities and should
strictly hold only in the limit
$T\to 0$. Given these constraints, 
the agreement between  the simple formula (\ref{MD.2}) for $\overline
v$ and the simulation results in Fig.~\ref{fig:vPhonon} is fairly satisfactory.

From the above discussion and using (\ref{5.0})  we obtain
\begin{equation}\label{MD.3}
\Lambda\cong 0.38\, \delta^{-2}T^{-2}
\end{equation}
for small $\delta$, $T$. From the measured values of $\kappa,\overline
v$ (see Figs.~\ref{fig:vPhonon} and \ref{fig:dDep}), 
$\Lambda$ can be determined according to (\ref{MD.1}). $\Lambda$ ranges
then from $\Lambda\sim1$ at $T=4$, 
$\delta=0.3$ to $\Lambda\sim4000$ at $T=0.1$, $\delta=0.08$. In the
simulations the system size is varied up to a few thousand depending on
the parameters.  Therefore, in our simulations, we have been able to
achieve $N\gtrsim\Lambda$ in the parameter range probed here. When
$\Lambda$ and $N$ are of the same order, it is not clear a priori if
the bulk limit has been reached.  In the subsequent section, we
provide an argument that we might still be able to estimate the
conductivity in such cases,  even when $N<\Lambda$. 

\begin{figure}
  \centering
  \includegraphics*[width=0.8\textwidth]{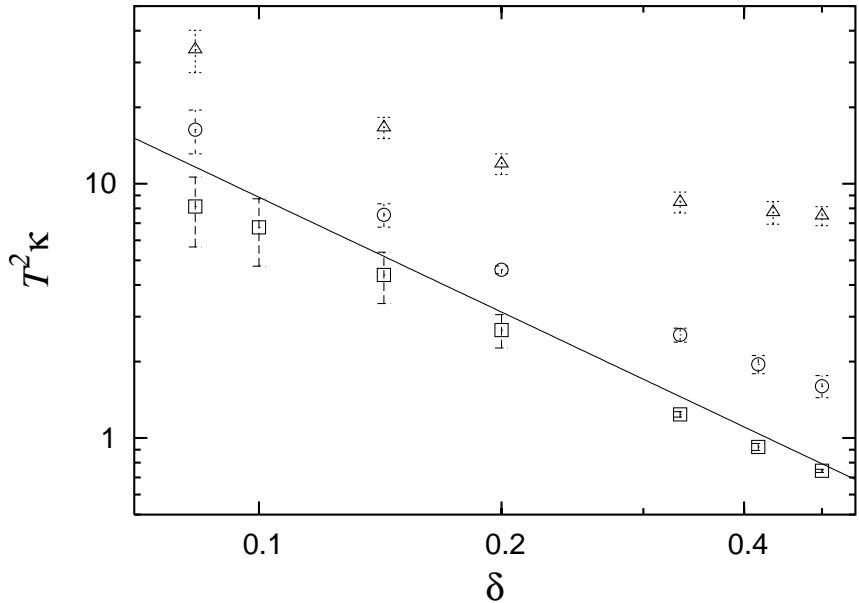}
  \caption{$T^2\kappa(T,1/\sqrt\delta,\delta,1)$ compared against the
  kinetic theory computation, $0.28\delta^{-3/2}$ (straight line).
  The data points are for $T=0.1\ (\Box), \ 0.4\ (\bigcirc),\ 4
  (\bigtriangleup)$.}
  \label{fig:dDep}
\end{figure}

One distinctive qualitative feature of kinetic theory is the leading
$\delta^{-3/2}$ dependence of
$T^2\kappa(T)$, see (\ref{5.0}).
To compare the molecular dynamics simulation results with kinetic
theory, we need to keep in mind that the agreement should hold
only when $T$ is small and the power law becomes exact when $\delta \to 0$.
In Fig. \ref{fig:dDep} the comparison is made
and we note a surprisingly
good agreement, which improves as the temperature is lowered, 
as to be expected. 


\section{Steady state temperature profile}\label{sec.5}
 \setcounter{equation}{0}

A second quantity of physical interest is the average
temperature profile. If difference in the boundary temperatures,
$\Delta T=T_{-}-T_{+}$, is small, then the profile is approximately linear. For
larger $\Delta T$ the temperature dependence of $\kappa$ will be
seen. As $\Delta T$ is further increased,  local
equilibrium will break down eventually  \cite{AK03}. We begin with 
working out the  temperature profile as predicted by the transport equation.

 The space-time dependent version of the Boltzmann equation
 (\ref{3.17}) reads
\begin{equation}\label{5.1}
\frac{\partial}{\partial t}W(x,k,t)+\frac{1}{2\pi}\omega'(k)
\frac{\partial}{\partial
x}W(x,k,t)=\mathcal{C}(W(x,\cdot,t))(k)\,,
\end{equation}
$\omega'(k)=\frac{d}{dk}\omega(k)$, where the collision operator
is local in $(x,t)$, i.e., it acts only on the wave number $k$.
 Under (\ref{5.1}) the phonon number
density
\begin{equation}\label{5.2}
  n(x,t)=\int_\mathbb{T}W(x,k,t)dk
\end{equation}
and the energy density
\begin{equation}\label{5.3}
  e(x,t)=\int_\mathbb{T}\omega(k)W(x,k,t)dk
\end{equation}
are locally conserved. The former one we regard as spurious, since
it has no analogue on the microscopic level.

Following the standard hydrodynamic scheme, since there are no convective
terms, the long time behavior
of (\ref{5.1}) is thus dictated by the solution of the coupled
nonlinear diffusion equations
\begin{equation}\label{5.4}
\frac{\partial}{\partial t}\binom{n}{e}=\frac{\partial}{\partial
x}D(\alpha,\beta)\frac{\partial}{\partial
x}\binom{\alpha}{\beta}\,.
\end{equation}
Here $\alpha$, $\beta$ are ``chemical potentials'' labelling the
stationary solutions, $W_{\alpha,\beta}(k)$, of (\ref{5.1}) as
\begin{equation}\label{5.5}
W_{\alpha,\beta}(k)=(\beta\omega(k)+\alpha)^{-1}
\end{equation}
for $(\alpha,\beta)\in\mathcal{D}$, where
$\mathcal{D}=\{\alpha,\beta|\omega(0)\beta>-\alpha$ for $\beta\geq
0$ and $\omega(\frac{1}{2})\beta>-\alpha$ for $\beta\leq 0\}$.
Then $\mathcal{D}\ni(\alpha,\beta)\mapsto(n,e)$ with
\begin{equation}\label{5.6}
n(\alpha,\beta)=\int_\mathbb{T}W_{\alpha,\beta}(k)dk\,,\quad
e(\alpha,\beta)=\int_\mathbb{T}\omega(k)W_{\alpha,\beta}(k)dk\,.
\end{equation}
In (\ref{5.4}) we insert the inverse function as defined on
$(\mathbb{R}_+)^2$, which is uniquely specified because of  convexity.
Secondly, $D(\alpha,\beta)$ is the $2\times 2$ matrix of Onsager
coefficients as given through a Green-Kubo formula analogous to
(\ref{3.22}). Following in spirit the arguments from \cite{S05} one obtains
\begin{equation}\label{5.7}
D(\alpha,\beta)=(2\pi)^{-2}\langle\omega'
(W_{\alpha,\beta})^2\binom{1}{\omega}\,,\,\frac{1}{\widetilde{L}}\,
(1\;\;\omega)\omega'(W_{\alpha,\beta})^2 \rangle
\end{equation}
with the linearized collision operator
\begin{eqnarray}\label{5.8}
&&\hspace{-26pt} \widetilde{L }f(k)= \frac{9\pi}{4}
\int_{\mathbb{T}^{3}}dk_1dk_2dk_3(\omega\omega_1\omega_2\omega_3)^{-1}
\delta(\omega+\omega_1-\omega_2-\omega_3)\nonumber\\
&&\hspace{23pt}\delta(k+k_1-k_2-k_3)
W_{\alpha,\beta}(k)W_{\alpha,\beta}(k_1)W_{\alpha,\beta}(k_2)
W_{\alpha,\beta}(k_3)\nonumber\\
&&\hspace{23pt} \big(f(k)+f(k_1)-f(k_2)-f(k_3)\big)\,.
\end{eqnarray}

\begin{figure}
  \centering
  \includegraphics*[width=0.8\textwidth]{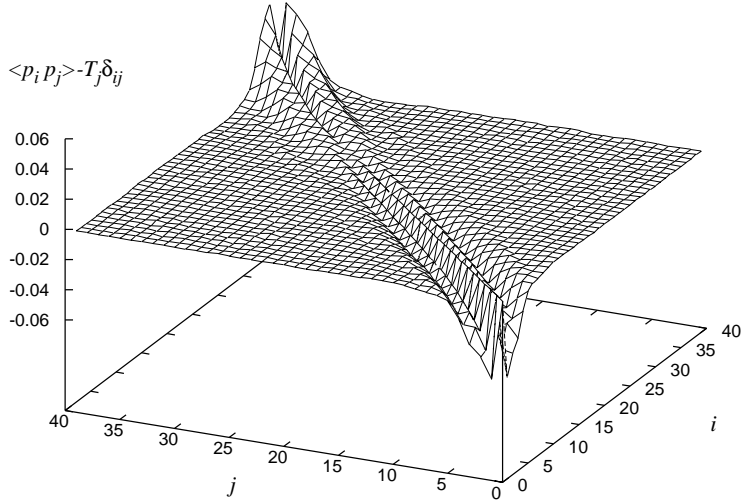}
  \caption{Steady state correlations $\langle p_i
  p_j\rangle-T_j\delta_{ij}$ for boundary 
    temperatures $T_{-} = 0.8$,  $T_{+} =1.2$, $\delta=1/3$, and $N=40$. }
  \label{fig:pp}
\end{figure}

At $\alpha=0$, $D(0,\beta)=D_\mathrm{ee}$ is independent of
$\beta$, which results in an important simplification. Let us
consider the steady state problem for (\ref{5.4}) in the slab
$[0,1]$ with boundary conditions $\alpha(0)=0$, $\alpha(1)=0$,
$\beta(0)=\beta_{-}$, $\beta(1)=\beta_{+}$. Then the solution is
given by
\begin{equation}\label{5.9}
\alpha(x)=0\,,\quad \beta(x)=\beta_{-}(1-x)+\beta_{+} x\,,\quad
0\leq x\leq 1\,.
\end{equation}
In particular the steady state energy flux is
\begin{equation}\label{5.10}
  j_\mathrm{e}=D_{\mathrm{ee}}(\beta_{+}-\beta_{-})\,.
\end{equation}
For a small temperature difference, $\beta_{-}=T^{-1}$,
$\beta_{+}=(T+\Delta T)^{-1}$, one has in approximation
\begin{equation}\label{5.10a}
  j_\mathrm{e}=-D_\mathrm{ee}T^{-2}\Delta T\,,
\end{equation}
where
$D_{\mathrm{ee}}=\langle\omega^{-2}g\,,L^{-1}\omega^{-2}g\rangle$
in agreement with (\ref{3.23}).

In molecular dynamics simulations the two ends of the chain, $j=1$
and $j=N$, are coupled to thermal reservoirs with $\Delta T/T$ of
order $0.4$ or less.  The local value of
$\alpha$ is extracted from the simulation. A simple test is provided by the
momentum covariance. If locally the Wigner function has the form
$W_{\alpha,\beta}(k)$, then
\begin{equation}\label{5.11} \langle
p_ip_{j}\rangle=\int_\mathbb{T}(\beta\omega(k)+\alpha)^{-1}\omega(k)
\cos(2\pi (i - j) k)dk\,,
\end{equation}
which reduces to $\langle p_i p_j\rangle=\beta^{-1}\delta_{ij}$
for $\alpha=0$. Expanding (\ref{5.11}) in $\alpha,\delta$ results in
$\langle p_j
p_{j+1}\rangle/\langle p^2_j\rangle\simeq
-\alpha\delta/2$. Numerically,  the steady state momentum correlation
$\langle p_i p_j\rangle$  is indeed strongly peaked 
at $i=j$ and decays rapidly.  The simulations therefore indicate
that $\alpha$ is small compared to $\omega$ and hence one
can safely set $\alpha=0$.
An example for off-diagonal correlations obtained from molecular
dynamics simulations is shown in 
Fig.~\ref{fig:pp} and, in this case, $\left|\langle p_j 
p_{j+1}\rangle\right|/\langle p_j^2\rangle \sim0.02$.  
This quantity is close to but  not quite  zero. 
This may be due to  $\alpha$ being  small, but not
quite zero, or due to finite size corrections to  local equilibrium.

\begin{figure}
  \centering
  \includegraphics*[width=0.8\textwidth]{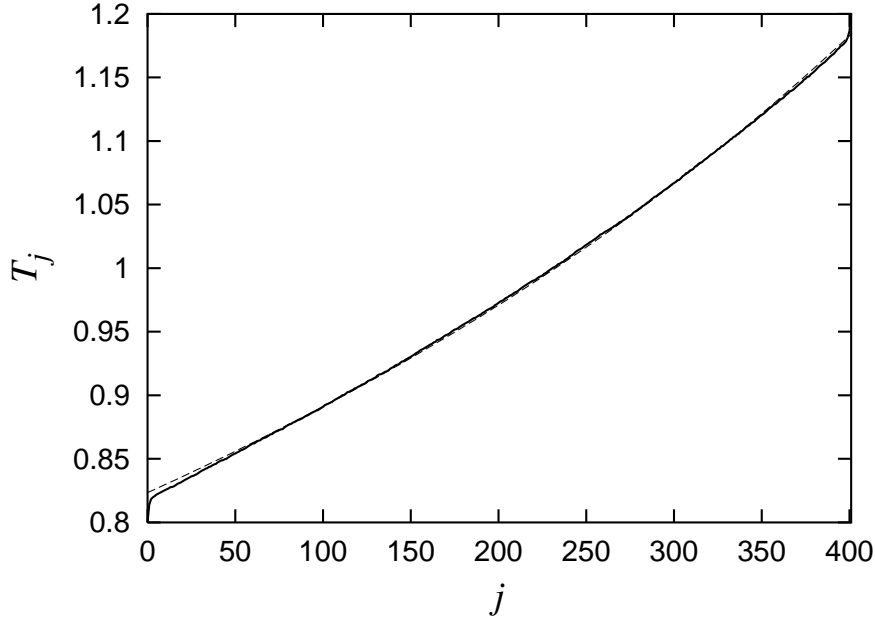}
  \caption{The temperature profile for $\delta=0.2$, boundary
    thermostat temperatures  $T_{-} = 0.8$, $T_{+} =1.2$, and $N=400$. 
The dashed line is a fit with a linear 
$1/T_j$-profile.}
  \label{fig:t-prof}
\end{figure}

In Fig.~\ref{fig:t-prof} we display a numerically generated steady state
profile. One notes that the profile lies slightly below the linear
interpolation. Indeed,  (\ref{5.9}) claims that $1/T_j$ is linear.
The dashed line is the corresponding fit. The excellent agreement 
is a further confirmation for $\alpha$ being small.

A generic temperature profile consists of three pieces: there are
two boundary jumps of equal size and concentrated over a few
lattice spacings and there is an, in essence, linear bulk piece. Only if
the effective temperature difference is large, one observes
deviations from linearity, as for example in Fig.~\ref{fig:t-prof}.
According to \cite{RLL67},
and as easily extended to the case under study, in
the harmonic limit
 boundary jumps would be exponentially localized
and the bulk piece would be flat. The observed temperature profile deviates
significantly from this harmonic limit for
all data points from Fig.~\ref{fig:dDep} in always having a nonzero slope.

The equal size boundary jumps are easily understood in the
context of Langevin reservoirs. The leftmost
particle, $j=1$, is then governed by the Langevin equation
\begin{eqnarray}\label{5.12}
&&\hspace{-26pt}\dot{q}_1=p_1\,,\nonumber\\
&&\hspace{-26pt}\dot{p}_1=-\omega^2_0q_1+\delta\omega^2_0q_2-\lambda
q^3_1-\gamma p_1+(2\gamma T_{-})^{1/2}\dot{b}(t)\,,
\end{eqnarray}
where $\gamma>0$ is the friction coefficient and $\dot{b}(t)$
white noise such that
$\langle\dot{b}(t)\dot{b}(t')\rangle=\delta(t-t')$. A corresponding
equation holds for the  
rightmost particle, $j=N$, with $T_{-}$ replaced by 
$T_{+}$. In the steady
state $\langle\dot{H}_1\rangle=0$, which implies
\begin{equation}\label{5.13}
\gamma(T_{-}-T_1)=J(N)\,,
\end{equation}
where $J(N)$ is the steady state current for chain length $N$. Similarly
\begin{equation}\label{5.14}
\gamma(T_{N}-T_+)=J(N)\,.
\end{equation}
In particular, the boundary jumps are equal.

To understand the full structure of the steady state one may
resort to (\ref{5.1}) restricted to the slab $[0,\ell]$. Then the steady
state Wigner function,  
$W(x,k)$, satisfies $\partial W/\partial t=0$ with the following boundary
conditions at the two ends, $x=0,\ell$,
\begin{eqnarray}\label{5.15}
&&\hspace{-26pt}W(0,k)=\frac{1}{2\pi}\omega'(k)T_{-}/\omega(k)
\quad\textrm{for}\;k>0\,,\nonumber\\
&&\hspace{-27pt}W(\ell,k)=-\frac{1}{2\pi}\omega'(k)T_{+}/\omega(k)
\quad\textrm{for}\;k<0\,,
\end{eqnarray}
where we assumed thermal sources together with complete
absorption. To obtain the steady state profile on
this basis would need considerable numerical effort. Therefore we
turn to a very much simplified model, which however retains the
gross features of the steady state. 

Energy is transported to the
right with velocity $+1$ and to the left with velocity $-1$.
Locally the velocity may switch its orientation  randomly with rate
$\eta$. Then the mean free path is $1/\eta$ and, as we will see, Fourier's law
holds with conductivity $\kappa=1/(2\eta)$. In the steady state the
average energy density satisfies
\begin{eqnarray}\label{5.16}
&&\hspace{-26pt}\frac{\partial}{\partial x}f_+(x)=
\eta\big(f_-(x)-f_+(x)\big) \,,\nonumber\\
&&\hspace{-34pt}-\frac{\partial}{\partial x}f_-(x)=
\eta\big(f_+(x)-f_-(x)\big)\,.
\end{eqnarray}
The local energy is $ f_+(x)+f_-(x) = T(x)$, which we identify with the 
local temperature. The energy current is then $ f_{+}(x)-f_{-}(x)$.
Energy is injected at $x=\ell$ and absorbed at $x=0$, i.e.
\begin{equation}\label{5.17}
f_-(\ell)=1\,,\quad f_+(0)=0\,,
\end{equation}
where partial absorption could readily be included.  Comparing to the case
$f_+(0)=1$
and $f_-(\ell)=1$, one concludes that the imposed right boundary temperature 
is $T_+ = 2$ and correspondingly the  imposed left boundary temperature 
$T_- = 0$. The solution
to (\ref{5.16}), (\ref{5.17}) reads
\begin{equation}\label{5.18}
f_+(x)=(1+\eta \ell)^{-1}\eta x\,,\quad f_-(x)=(1+\eta
\ell)^{-1}(\eta x+1)\,,
\end{equation}
which yields the temperature profile
\begin{equation}\label{5.19}
T(x)= f_+(x)+f_-(x)=(1+\eta \ell)^{-1}(2 \eta
x+1)
\end{equation}
and the steady state current $J(\ell)$ for slab length $\ell$ as
\begin{equation}\label{5.20}
J(\ell)=f_+(x)-f_-(x)=(1+\eta \ell)^{-1}\,.
\end{equation}
Thus at both ends the boundary jump equals $(1+\eta
\ell)^{-1}$ and the effective temperature difference is $\Delta
T=(1+\eta \ell)^{-1}2\eta \ell$. Hence
\begin{equation}\label{5.21}
J(\ell)= -\kappa\Delta T/\ell\,.
\end{equation}
Thus, even if $\ell \ll \eta^{-1}$, the correct bulk current is
extracted through (\ref{5.21}). Of course, as for fixed $\ell$ the
mean free path increases, so does the relaxation time and longer
simulation times would be needed.

To come back to the molecular dynamics computation,  the simulation
time is sufficiently long so that the steady state is reached.
For most of the data presented, we obtain the conductivity from
lattices with $N\gg \Lambda$ and are confident that they should be
reliable. At $T=0.1$ data with small $\delta$ are computed for lattices
with $N \sim \Lambda$. Given the reasoning within the simplified model and
the tendency of the data for larger $T$, we believe that a reasonable
estimate for the conductivities has been obtained even for these
cases.


\section{Conclusions}\label{sec.7}
\setcounter{equation}{0}
The Boltzmann-Peierls equation retains the exact dispersion relation
and the type of nonlinearity of the Hamiltonian model. For example, had we
considered a cubic nonlinearity, then Eq.~(\ref{3.17}) would be
quadratic in $W$. For a nonlinearity which depends on the nearest neighbor
relative displacements there would appear the additional factor
$|\prod^4_{j=1}(1-\exp(i2\pi k_j))|^2$ in the collision rate. It is
remarkable that with this input the qualitative features of the
``low'' temperature thermal conductivity are so well recovered.

It seems to us that the Boltzmann-Peierls equation has never been
tested in comparable precision before. The peak time for the
experimental investigation of phonon thermal conductivity was in the
late 50'ies and early 60'ies. A quantitative comparison with the
theory was hampered from two sides: (1) The dispersion relation and
the anharmonicities of a given dielectric crystal are not so readily
available. (2) One needs considerable numerical effort to reliably
obtain the thermal conductivity from the transport equation. Thus
mostly one had to be satisfied with qualitative predictions, as for
example the $1/T$-dependence of the thermal conductivity in the 
presence of only three-phonon collisions \cite{Zi}. More recently, 
molecular dynamics simulations have become
available, for example see \cite{MK04} and references therein. 
Compared to these more material science oriented studies, we achieve 
a much larger system size,
due to one instead of three spatial dimensions,
and we test the simulation data directly against the transport equation 
without further approximations.
\bigskip\\
{\bf Acknowledgments.} 
KA  was supported in part by Grant--in--Aid for Scientific Research
from the Ministry of Education, Culture, Sports, Science and
Technology of Japan.
JL and HS thank Jean Bricmont, Antti Kupiainen,
Raphael Lefevere, and Alain Schenkel for most instructive discussions,
which merged into the present study.


\begin{appendix}
\section{Appendix}\label{sec.A}
 \setcounter{equation}{0}

 We follow Sect.~11 of \cite{S05}. The initial measure is
 translation invariant and uniquely characterized by the
 covariance of (\ref{3.12}). We want to establish that
\begin{equation}\label{A.1}
W^\lambda(k,\lambda^{-2}t)=W(k)+t\mathcal{C}(W)(k)+\mathcal{O}(t^2)
\end{equation}
for small $\lambda$.

By (\ref{3.7}) the vertex weight is given by
\begin{equation}\label{A.2}
\phi(k,k_1,k_2,k_3)=(16\omega\omega_1\omega_2\omega_3)^{1/2}\,.
\end{equation}
We use the expansion through Feynman diagrams as obtained from the
iteration of (11.7) in \cite{S05}. Since the Hamiltonian has a
quartic nonlinearity, each interaction leads to a branching into 3
lines, compare with Fig.~\ref{fig:dDep} of \cite{S05}.

To order $\lambda^0$ we simply have $\delta(q-p)W(q)$. The order
$\lambda$ term is purely imaginary, thus vanishes, because to
every diagram there is its complex conjugate, denoted by
\textit{c.c.}. Thus we are left with the order $\lambda^2$. It has
8 ways of branching. For a given branching there are 15 Gaussian
pairings and 8 possible orientations of the internal lines, which
in total amounts to 960 diagrams. They will be divided into
subleading and
leading.\smallskip\\
(i) There are 144 diagrams of the type\\[1em]
\begin{minipage}{\textwidth}
\setlength{\unitlength}{1cm}
\begin{picture}
(9,3.2)(-3.5,0)  \put(7.5,-0.1){$0$} \put(7.5,0.9){$t_1$}
\put(7.5,1.9){$t_2$} \put(7.5,2.9){$t$}
\put(1.4,2.5){$q$}\put(5.4,2.5){$p$} \linethickness{0.1pt}
\put(0,0){\line(1,0){7}} \put(0,1){\line(1,0){7}}
\put(0,2){\line(1,0){7}} \put(0,3){\line(1,0){7}}
\thicklines
\put(1,-0.1){\line(0,1){2.1}} \put(2,-0.1){\line(0,1){3.1}}
\put(3,-0.1){\line(0,1){2.1}} \put(4,-0.1){\line(0,1){1.1}}
\put(5,-0.1){\line(0,1){3.1}} \put(6,-0.1){\line(0,1){1.1}}
\put(1,2){\line(1,0){2}} \put(4,1){\line(1,0){2}}
\put(1,-0.1){\line(1,0){1}} \put(3,-0.1){\line(1,0){1}}
\put(5,-0.1){\line(1,0){1}} \put(5,2.5){\vector(0,1){0.15}}
 \put(2,2.5){\vector(0,-1){0.15}}
\put(2,2){\circle*{0.17}} \put(5,1){\circle*{0.17}}
\end{picture}
\end{minipage}\\[1.5 em]
We set $\tau=t_2-t_1$. Their sum is then
\begin{equation}\label{A.3}
36\sum_{\sigma_2=\pm 1}\delta(q-p)\int_{\mathbb{T}^2} dk_1 dk_3
\phi(q,k_1,q,k_3)^2 W_1W(-\sigma_2q)
W_3(e^{i\tau\omega(q)(1+\sigma_2)}+c.c.)\,.\smallskip
\end{equation}
(ii) There are 144 diagrams of the type\\[1em]
\begin{minipage}{\textwidth}
\setlength{\unitlength}{1cm}
\begin{picture}
(9,3.2)(-3.5,0)  \put(7.5,-0.1){$0$} \put(7.5,0.9){$t_1$}
\put(7.5,1.9){$t_2$} \put(7.5,2.9){$t$}
\put(1.4,2.5){$q$}\put(6.4,2.5){$p$} \linethickness{0.1pt}
\put(0,0){\line(1,0){7}} \put(0,1){\line(1,0){7}}
\put(0,2){\line(1,0){7}} \put(0,3){\line(1,0){7}}
\thicklines
\put(1,-0.1){\line(0,1){2.1}} \put(2,-0.1){\line(0,1){3.1}}
\put(3,-0.1){\line(0,1){1.1}} \put(4,-0.1){\line(0,1){2.1}}
\put(5,-0.1){\line(0,1){1.1}} \put(6,-0.1){\line(0,1){3.1}}
\put(1,2){\line(1,0){3}} \put(3,1){\line(1,0){2}}
\put(1,-0.1){\line(1,0){1}} \put(3,-0.1){\line(1,0){1}}
\put(5,-0.1){\line(1,0){1}} \put(6,2.5){\vector(0,1){0.15}}
 \put(2,2.5){\vector(0,-1){0.15}}
\put(2,2){\circle*{0.17}} \put(4,1){\circle*{0.17}}
\end{picture}
\end{minipage}\\[1.5 em]
Their sum is
\begin{equation}\label{A.4}
36\sum_{\sigma_2=\pm 1}\delta(q-p)\int_{\mathbb{T}^2} dk_1 dk_3
\phi(q,k_1,q,k_3)^2 W_1W(q)W_3\sigma_2
(e^{i\tau\omega(q)(1+\sigma_2)}+c.c.)\,. \smallskip
\end{equation}
(iii) There are 288 diagrams of the type\\[1.5 em]
\begin{minipage}{\textwidth}
\setlength{\unitlength}{1cm}
\begin{picture}
(9,3.2)(-3.5,0)  \put(7.5,-0.1){$0$} \put(7.5,0.9){$t_1$}
\put(7.5,1.9){$t_2$} \put(7.5,2.9){$t$}
\put(1.4,2.5){$q$}\put(6.4,2.5){$p$} \linethickness{0.1pt}
\put(0,0){\line(1,0){7}} \put(0,1){\line(1,0){7}}
\put(0,2){\line(1,0){7}} \put(0,3){\line(1,0){7}}
\thicklines
\put(1,-0.2){\line(0,1){2.2}} \put(2,-0.1){\line(0,1){3.1}}
\put(3,-0.1){\line(0,1){1.1}} \put(4,-0.1){\line(0,1){2.1}}
\put(5,-0.1){\line(0,1){1.1}} \put(6,-0.2){\line(0,1){3.2}}
\put(1,2){\line(1,0){3}} \put(3,1){\line(1,0){2}}
\put(1,-0.2){\line(1,0){5}} \put(2,-0.1){\line(1,0){1}}
\put(4,-0.1){\line(1,0){1}} \put(6,2.5){\vector(0,1){0.15}}
 \put(2,2.5){\vector(0,-1){0.15}}
\put(2,2){\circle*{0.17}} \put(4,1){\circle*{0.17}}
\end{picture}
\end{minipage}\\[1.5 em]
Their sum is
\begin{equation}\label{A.5}
36\sum_{\sigma_1,\sigma_2=\pm 1}\delta(q-p)\int_{\mathbb{T}^2}
dk_1 dk_3 \phi(q,k_1,q,k_3)^2 W_1W(q)W_3\sigma_2
(e^{i\tau\omega_1(\sigma_1+\sigma_2)}+c.c.)=0\,.
\end{equation}

In (i) and (ii) the terms independent of $\tau$ cancel each other.
The remaining terms are proportional to $\cos(2\omega(q)\tau)$ and
thus of order $\lambda^2$ after time-integration. We are left with
384 leading diagrams.\smallskip\\
(iv) The gain term results from 96 diagrams of the type\\[1em]
\begin{minipage}{\textwidth}
\setlength{\unitlength}{1cm}
\begin{picture}
(9,3.2)(-3.5,0)  \put(7.5,-0.1){$0$} \put(7.5,0.9){$t_1$}
\put(7.5,1.9){$t_2$} \put(7.5,2.9){$t$}
\put(1.4,2.5){$q$}\put(5.4,2.5){$p$} \linethickness{0.1pt}
\put(0,0){\line(1,0){7}} \put(0,1){\line(1,0){7}}
\put(0,2){\line(1,0){7}} \put(0,3){\line(1,0){7}}
\thicklines
\put(1,-0.3){\line(0,1){2.3}} \put(2,-0.2){\line(0,1){3.2}}
\put(3,-0.1){\line(0,1){2.1}} \put(4,-0.1){\line(0,1){1.1}}
\put(5,-0.2){\line(0,1){3.2}} \put(6,-0.3){\line(0,1){1.3}}
\put(1,2){\line(1,0){2}} \put(4,1){\line(1,0){2}}
\put(1,-0.3){\line(1,0){5}} \put(2,-0.2){\line(1,0){3}}
\put(3,-0.1){\line(1,0){1}} \put(5,2.5){\vector(0,1){0.15}}
 \put(2,2.5){\vector(0,-1){0.15}}
\put(2,2){\circle*{0.17}} \put(5,1){\circle*{0.17}}
\end{picture}
\end{minipage}\\[1.5 em]
Their sum is
\begin{eqnarray}\label{A.6}
&&\hspace{-20pt} 6\sum_{\sigma_1,\sigma_2,\sigma_3=\pm 1}
\delta(q-p)\int_{\mathbb{T}^3} dk_1 dk_2 dk_3
\phi(q,k_1,k_2,k_3)^2 \delta(q+\sigma_1k_1+\sigma_2k_2+\sigma_3k_3)
\nonumber\\
&&\hspace{60pt} W_1W_2W_3(e^{i\tau(\omega(q)+
\sigma_1\omega_1+\sigma_2\omega_2+\sigma_3\omega_3)}+c.c.)\,.
\smallskip
\end{eqnarray}
(v) The loss term  results from 288 diagrams of the type
\\[1em]
\begin{minipage}{\textwidth}
\setlength{\unitlength}{1cm}
\begin{picture}
(9,3.2)(-3.5,0)  \put(7.5,-0.1){$0$} \put(7.5,0.9){$t_1$}
\put(7.5,1.9){$t_2$} \put(7.5,2.9){$t$}
\put(1.4,2.5){$q$}\put(6.4,2.5){$p$} \linethickness{0.1pt}
\put(0,0){\line(1,0){7}} \put(0,1){\line(1,0){7}}
\put(0,2){\line(1,0){7}} \put(0,3){\line(1,0){7}}
\thicklines
\put(1,-0.2){\line(0,1){2.2}} \put(2,-0.1){\line(0,1){3.1}}
\put(3,-0.1){\line(0,1){1.1}} \put(4,-0.2){\line(0,1){2.2}}
\put(5,-0.1){\line(0,1){1.1}} \put(6,-0.1){\line(0,1){3.1}}
\put(1,2){\line(1,0){3}} \put(3,1){\line(1,0){2}}
\put(1,-0.2){\line(1,0){3}} \put(2,-0.1){\line(1,0){1}}
\put(5,-0.1){\line(1,0){1}} \put(6,2.5){\vector(0,1){0.15}}
 \put(2,2.5){\vector(0,-1){0.15}}
\put(2,2){\circle*{0.17}} \put(4,1){\circle*{0.17}}
\end{picture}
\end{minipage}\\[1.5 em]
Their sum is
\begin{eqnarray}\label{A.7}
&&\hspace{-30pt} 6\sum_{\sigma_1,\sigma_2,\sigma_3=\pm 1}
\delta(q-p)\int_{\mathbb{T}^3} dk_1 dk_2 dk_3
\phi(q,k_1,k_2,k_3)^2 \delta(q+\sigma_1k_1+\sigma_2k_2+\sigma_3k_3)
\nonumber\\
&&\hspace{0pt} W(q)(\sigma_1
W_2W_3+\sigma_2W_1W_3+\sigma_3W_1W_2)(e^{i\tau(\omega(q)+
\sigma_1\omega_1+\sigma_2\omega_2+\sigma_3\omega_3)}+c.c.)\,.
\end{eqnarray}

In (iv) and (v) we use that
\begin{equation}\label{A.8}
\lim_{\lambda\to 0}\lambda^2\int^{\lambda^{-2}t}_0 dt_2
\int^{t_2}_0 dt_1(e^{i\omega(t_2-t_1)}+c.c.)=2\pi
t\delta(\omega)\,,
\end{equation}
when integrated against a smooth, rapidly decreasing test
function. By adding (iv) and (v) one obtains the collision
operator from Eq.~(\ref{3.15}) with the prefactor $12\pi$.

\end{appendix}


\end{document}